\documentclass{jpsj2}

\newcommand{\rvec}{\mbox{\boldmath $r$}}

\newcommand{\tldetotal}{\widetilde{E}_{\mbox{\scriptsize total}}}
\newcommand{\etotal}{E_{\mbox{\scriptsize total}}}
\newcommand{\RLR}{R_{\mbox{\tiny LR}}}
\newcommand{\RKRNL}{R_{\mbox{\tiny KR}}}

\begin{document}

\title{Augmented orbital minimization method for linear scaling 
electronic structure calculations}

\author{E\hspace{0.4mm}i\hspace{0.4mm}j\hspace{0.4mm}i  Tsuchida}

\inst{Research Institute for Computational Sciences, 
   Natl Inst of AIST, 
   Tsukuba Central 2, Umezono 1-1-1, Tsukuba 305-8568}

\abst{
We present a novel algorithm which can overcome the drawbacks 
of the conventional linear scaling method 
with minimal computational overhead. 
This is achieved by introducing additional constraints, thus 
eliminating the redundancy of the orbitals. 
The performance of our algorithm is evaluated in 
ab initio molecular-dynamics simulations as well as in a model system. 
}

\kword{density-functional theory, finite-element method, 
linear scaling method, orbital minimization, electronic structure}

\maketitle

\newpage

\section{INTRODUCTION}

Electronic structure calculations often provide very accurate 
physical and chemical properties of various microscopic systems 
from first principles, 
which makes them attractive to experimentalists 
as well as theoreticians \cite{HK,KS,CP,DFTREV1,DFTREV2}. 
However, the computational cost of such calculations 
grows cubically (or faster) with the number of atoms in the system, 
thereby limiting the maximum number of atoms to $\approx 10^3$ 
on today's computers. 
Therefore, much effort has been devoted to the development of 
so-called linear scaling methods, 
whose computational cost grows only linearly 
with system size \cite{DFTREV2,GOED99,SCU99,WU02},  
usually by making some assumptions 
about the electronic structure of the system \cite{KOHN96}. 

The emergence of the linear scaling methods has also 
promoted the development of various discretization schemes in real space 
in the last decade \cite{BECK00,PKFS05,TORS06,RSBOOK}, such as 
finite difference and finite element methods. 
These real space methods are considered 
more appropriate for linear scaling methods 
than plane waves, because they can easily 
take advantage of the localization of electrons \cite{KOHN96} 
while retaining systematic convergence. 
Alternatively, the use of atomic basis set 
in linear scaling methods is also an attractive approach 
\cite{YANG91,SCU99,FMO99,OZ04}. 

From the point of view of computational cost, 
the orbital minimization method (OMM) \cite{GP92,MG94,KMG95,ORD95}, 
which is designed for nonmetallic systems, 
is among the most promising linear scaling algorithms proposed so far. 
Moreover, OMM is easy to implement, and is able to deal with 
nonorthogonal basis functions without much difficulty. 
Therefore, much work has been carried out on the implementation of OMM, 
including first-principles calculations using 
real-space methods \cite{FB00,FG04,FG06,FEM2,RACZ01,MORT01,SHIM01}. 
Unfortunately, if localization constraints are imposed 
on the orbitals to achieve linear scaling, 
a naive implementation of OMM suffers from 
several drawbacks \cite{GOED99}, 
which has discouraged the use of OMM 
in realistic applications to date. 
In the present paper, we propose a simple yet effective algorithm 
which can overcome the drawbacks of OMM 
when the electronic structure of the system is 
qualitatively predictable. 

\newpage

\section{ALGORITHM}
First of all, we briefly describe the basic formalism of 
electronic structure calculations here. 
For notational simplicity, only non-self-consistent 
problems are considered here, but extention to 
self-consistent ones is straightforward. 
Moreover, we assume that the orbitals are real, 
and that there is no spin degeneracy. 
We also assume the presence of an energy gap 
between the occupied and unoccupied states throughout the paper. 
Then, the conventional total energy functional $\tldetotal$ is 
given by 
\begin{equation}
\tldetotal [\widetilde{\psi}] = \sum_{i=1}^N \widetilde{H}_{ii}, 
\end{equation}
where the Hamiltonian ${\cal H} = -\nabla^2 + {\cal V}$, 
$\widetilde{H}_{ii} = <\!\widetilde{\psi}_i \,|\, {\cal H} \, 
| \,\widetilde{\psi}_i\!>$, 
${\cal V}$ is the potential felt by the electrons, and 
$N$ is the number of occupied orbitals \cite{FNOTE7}. 
If $\tldetotal$ is minimized 
with repect to the orbitals 
$\{\widetilde{\psi}_i (\rvec)\}_{i=1}^{N}$ 
under the orthonormality constraints 
\begin{equation}
\label{ORTHOG}
<\!\widetilde{\psi}_i \, | \, \widetilde{\psi}_j\!> = \delta_{ij}, 
\end{equation}
$\tldetotal$ and $\{{\widetilde{\psi}_i}\}$ 
will converge to the sum of the $N$ lowest eigenvalues of 
${\cal H}$ and corresponding eigenstates, respectively, 
except for the degrees of freedom associated with 
any unitary transformation. 
This redundancy can be exploited to construct 
the maximally localized Wannier functions (MLWFs) \cite{MLWF97,BRGHLD00}, 
whose spread in real space, 
$$\Omega = \sum_{i=1}^N (<\!r^2\!>_i - <\!r\!>_i^2), $$ 
is minimum among all states 
given by the unitary transformation of the eigenstates. 
An efficient calculation of MLWFs along the trajectory of 
Car-Parrinello dynamics is also an active area of research 
\cite{SHARMA03,THOM04,IFTIM04,KIRCH04}. 
In the following, MLWFs are denoted by $\{w_i(\rvec)\}_{i=1}^N$. 

On the other hand, the generalized 
total energy functional \cite{SCPB89,APJ,GP92,KV} used in OMM is given by  
\begin{equation}
\etotal [{\psi}] = \sum_{i,j=1}^N (S^{-1})_{ij} \cdot H_{ij}, 
\end{equation}
where $ H_{ij} = <\!\psi_i \,|\, {\cal H} \, | \,\psi_j\!>$, 
and the overlap matrix $S$ is defined by 
$ S_{ij} = <\!\psi_i \,|\, \psi_j\!> $. 
The minimum value of $\etotal$ agrees with 
that of $\tldetotal$, and $\{\psi_i\}$ that 
minimize $\etotal$ are nonorthogonal 
functions that span the same subspace as 
the $N$ lowest eigenstates of ${\cal H}$.
While there are several variants of 
this functional \cite{MG94,KMG95,ORD95} 
which rely on the Neumann expansion of $S^{-1}$, 
we will not go into detail here. 
In analogy with the case of $\tldetotal$, 
$\etotal$ is invariant under the linear transformation 
\begin{equation}
\label{XIJ}
|\,\psi_i\!> = \sum_j X_{ij} |\,\psi'_j\!>
\end{equation}
for any nonsingular matrix $X$ of size $N$. 
Therefore, attempts have been made to construct 
nonorthogonal localized orbitals (NOLOs), 
which can be more localized than MLWFs 
by taking advantage of the higher 
degree of freedom \cite{LIU00,FENG04,QMC05}. 
However, special attention has to be paid to the risk of 
falling into linearly dependent states 
while constructing NOLOs. 

\begin{figure}
  \begin{center}
  \includegraphics[width=80mm]{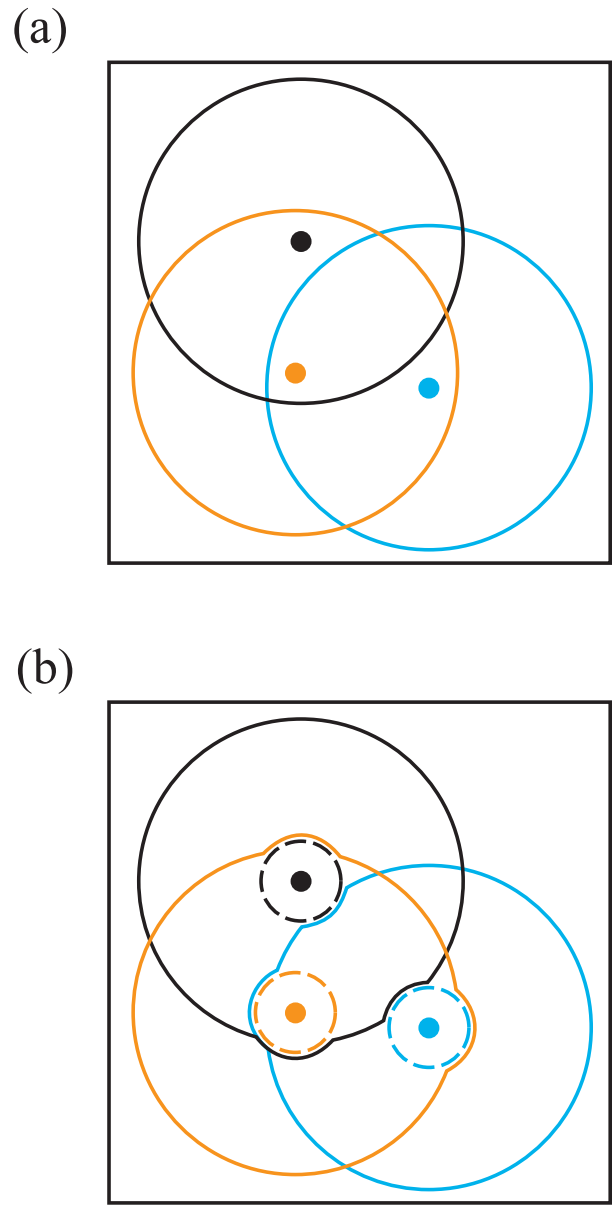}
  \end{center}
  \caption{(a) Conventional definition of the 
	localization regions in OMM (solid lines). 
	Filled circles denote the centers of localization, which are 
	usually either atomic positions or bond centers. 
	(b) Localization regions in AOMM. 
	Kernel regions are shown with dashed lines. }
  \label{LOCREG}
\end{figure}

In order to achieve linear scaling with the OMM, 
localization constraints are imposed 
on the orbitals (Fig.\ref{LOCREG} (a)) 
during the minimization of $\etotal$ \cite{GP92}. 
When each orbital is strictly localized within a given region of space, 
called the localization region (LR), 
$S$ and $H$ would be sparse matrices, 
and the computational cost of evaluating each nonzero element of 
$S$ and $H$ would be independent of system size \cite{FNOTE4}. 
Therefore, $\etotal$ as well as its gradient can be 
calculated with linear scaling 
in a straightforward manner \cite{FNOTE1}. 
Moreover, the optimized orbitals are expected to be 
good approximations to MLWFs, 
which are least likely to be influenced 
by the localization constraints among the unitary transformation 
of the ground state. 
Therefore, the centers of LRs are usually chosen 
as close to those of MLWFs as possible. 

Unfortunately, in the presence of localization constraints, 
iterative minimization of the total energy becomes extremely 
difficult \cite{GOED99}, 
often requiring hundreds or thousands of iterations to converge. 
Furthermore, the orbitals can be 
trapped at local minima during the minimization 
process \cite{GOED99,FG06,KMG95}, 
which results in poor conservation of the total energy 
in molecular-dynamics simulations. 
The major source of these problems is that 
$\etotal$ has a pathological shape around the minimum, 
which arises from the fact that 
$\etotal$ is only approximately invariant under the 
linear transformation of Eq.(\ref{XIJ}) 
in the presence of localization constraints \cite{GOED99,BOW00}. 

While much effort has been made to overcome 
these problems \cite{GOED99,KMG95,STP98}, 
the performance and reliability of OMM under the localization constraints 
still appear to be insufficient for routine use 
in realistic applications. 
In the following, we present a simple prescription to make 
OMM a practical linear scaling algorithm 
with minimal computational overhead. 
To this end, we introduce here the concept of kernel region (KR), 
which plays an important role in the algorithm explained below. 
For simplicity, we assume that only one orbital is assigned to 
each LR, but extension to the multi-orbital case is straightforward. 
Then, KRs of a given system are generated 
under the following conditions: 
\begin{enumerate}
\item[(a)] Each LR includes its own KR, which preferably includes 
	the center of a MLWF. 
\item[(b)] There is no overlap between any two KRs.
\item[(c)] No partial overlap between any LR and KR is allowed. 
\end{enumerate}
An example of a set of LRs and KRs that satisfy these conditions 
is shown in Fig.\ref{LOCREG} (b). 
In practice, we first generate LRs and KRs temporarily, 
e.g. by the distance criterion, that satisfy conditions (a) and (b). 
Then, if more(less) than a given fraction (say, 40 \%) of each KR is 
included in some other LR, the border of that LR is 
modified to include(exclude) the KR completely, thus satisfying 
condition (c). 
Since KRs are usually much smaller than LRs, 
these modifications will not 
have a significant impact on the shape of LRs. 
In the following, LRs and KRs that satisfy the above conditions are 
denoted by $\{L_i\}_{i=1}^N$ and $\{K_i\}_{i=1}^N$, respectively. 

We now define the kernel functions $\{\chi_i (\rvec)\}_{i=1}^N$ that 
have the following properties: 
\begin{enumerate}
\item[(i)] $\chi_i (\rvec)$ approximates $w_i(\rvec)$ when $\rvec \in K_i$.
\item[(ii)] $\chi_i (\rvec) = 0$ when $\rvec \notin K_i$.
\item[(iii)] $<\!\chi_i \, | \, \chi_i\!>  = 1$.
\end{enumerate}
Therefore, $<\!\chi_i \, | \, \chi_j\!>  = \delta_{ij}$ 
is satisfied automatically. 

In the augmented orbital minimization method (AOMM), 
$\etotal$ is minimized with respect to the 
localized orbitals $\{\psi_i\}$  under the additional constraints that 
\begin{equation}
\label{LORTHO}
<\!\chi_j \, | \, \psi_i\!> = 0 
\end{equation}
for any $j \ne i$, where $ i,j=1,2,\cdots N $. 
The role of these constraints is to 
orthogonalize $\psi_i(\rvec)$ approximately to $w_j(\rvec)$ for any 
$j \ne i$, in the hope that 
$\psi_i(\rvec)$ will be a good approximation to $w_i(\rvec)$ 
at the minimum. 
Eq.(\ref{LORTHO}) is satisfied by an explicit orthogonalization as
\begin{equation}
|\,\psi_i'\!> = \hat{P}_i |\,\psi_i\!> =  
|\,\psi_i\!> -  \sum_{j \ne i} | \, \chi_j\!> <\!\chi_j \, |\,\psi_i\!>, 
\end{equation}
where the projection operator is given by 
\begin{equation}
\hat{P}_i =I-\sum_{j \ne i} | \, \chi_j\!> <\!\chi_j \, |.  
\end{equation}
Summation with respect to $j$ should be taken 
only if $K_j \in L_i$, because $ <\!\chi_j \, | \, \psi_i\!> \equiv 0$ 
otherwise. 
Therefore, the computational cost of projection is 
relatively minor, scaling only linearly with system size. 
Note that if $\psi_i (\rvec)$ is localized within $L_i$, 
so is $\psi_i' (\rvec)$ due to the properties of KRs and kernel functions. 
Moreover, each orbital remains unchanged inside its own KR 
after the projection, i.e. 
$\psi_i'(\rvec)=\psi_i(\rvec)$ if $\rvec \in K_i$.  

There is no unique way to define the kernel functions for given KRs, 
but if those KRs are used as the LRs in the conventional OMM, 
the optimized orbitals will serve as the kernel functions. 
These are called static kernel functions, since they 
do not change during the electronic structure calculations. 
Note that there is no slow convergence or local minima problem  
when the LRs do not overlap. 
An alternative way to define the kernel functions is to 
use a mask function $m_i (\rvec)$, such that 
$m_i (\rvec)=1$ when $\rvec \in K_i$ and $m_i (\rvec)=0$ 
otherwise \cite{FNOTE2}. 
Then, if the orbitals are reasonably close to the ground state, 
$\{m_i (\rvec) \psi_i (\rvec)\}_{i=1}^N$ can be used as 
the kernel functions after normalization. 
We call them dynamic kernel functions, because they are 
updated at every step of the minimization. 

\begin{figure}
  \begin{center}
  \includegraphics[width=70mm]{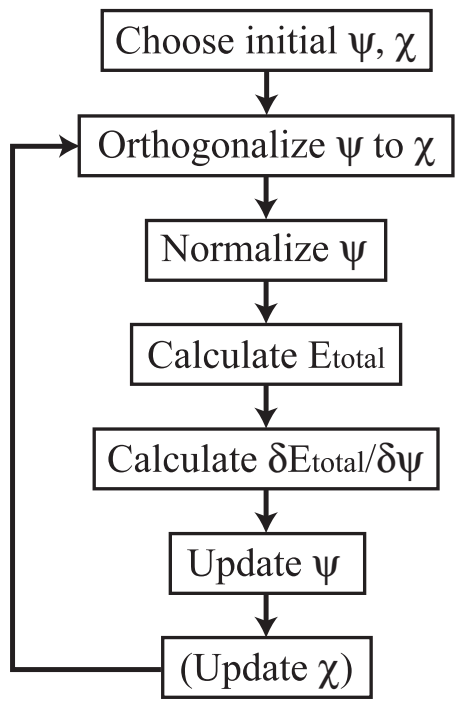}
  \end{center}
  \caption{Flow chart of AOMM for a given ionic configuration. 
	The loop is repeated until a convergence criterion is satisfied. 
	Update of $\chi$ can be skipped if static kernel functions are used.}
  \label{FLOWCHART}
\end{figure}

When the kernel functions do not depend on the orbitals, 
the gradient of $\etotal$ under the constraints of 
Eq.(\ref{LORTHO}) is given by 
\begin{equation}
|\,g_i\!> = \hat{P}_i \, \frac{\partial \etotal}{\partial \psi_i}, 
\end{equation}
which can also be evaluated with linear scaling effort. 
If dynamic kernel functions are used, 
a correction term is required to take into account 
the dependence of $\{\chi_i\}$ on $\{\psi_i\}$, 
which, however, can be calculated in a straightforward manner. 
Fig.\ref{FLOWCHART} shows the flow chart of the electronic structure 
calculation for a given ionic configuration in the AOMM.  

\newpage

\section{RESULTS}
\begin{figure}
  \begin{center}
  \includegraphics[width=90mm]{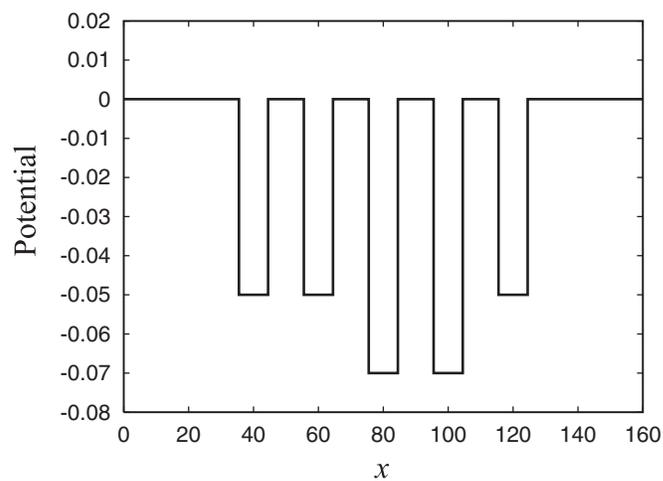}
  \end{center}
  \caption{Five square potential wells of widths 9 are centered at 
	40,60,80,100, and 120.}
  \label{POT1D}
\end{figure}

\begin{figure}
  \begin{center}
  \includegraphics[width=90mm]{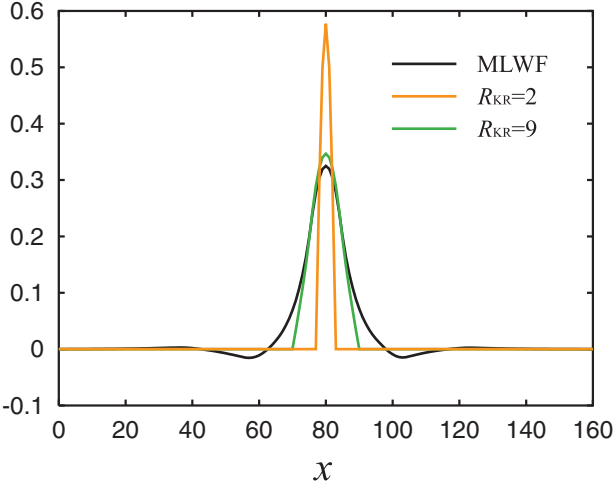}
  \end{center}
  \caption{MLWF and kernel functions which belong to the central KR.}
  \label{FIGKRNL}
\end{figure}

The performance of our algorithm is first evaluated 
in a simple one-dimensional problem. 
We consider a system consisting of 5 electrons 
in the potential wells shown in Fig.\ref{POT1D}, 
where $x=0,1, \cdots, 160$, and 
vanishing boundary conditions are imposed on the orbitals. 
When a 3-point finite-difference approximation is used 
for the Laplacian, 
the Hamiltonian ${\cal H}$ is given by a tridiagonal matrix of 
size 161$\times$161 as  
\begin{equation}
{\cal H}=
\left(
\begin{array}{ccccc}
2 + v_0 &   -1     &        &    &   \\
     -1 & 2 + v_1  & -1     &    &   \\
        & \ldots   & \ldots & \ldots       &   \\
        &          &     -1 &  2 + v_{159} & -1\\
        &          &        &           -1 & 2 + v_{160}\\
\end{array}
\right).
\end{equation}
\ \\
We used 5 pairs of LRs and KRs centered at 
40, 60, 80, 100, and 120, the radii of which are denoted by 
$\RLR$ and $\RKRNL$, respectively. 
Therefore, each LR(KR) is given 
2$\RLR$+1 (2$\RKRNL$+1) degrees of freedom. 
It is worth noting that in this system 
the conditions (a)-(c) given in the previous section translate into 
the inequalities as follows: 
(a) $\RKRNL \le \RLR$, (b) $0 \le \RKRNL \le 9$, 
and (c) $ 20-\RKRNL \le \RLR < 20+\RKRNL, 
40-\RKRNL \le \RLR < 40+\RKRNL, \cdots$ are not allowed. 
The centers of the MLWFs constructed from the 5 lowest eigenstates of 
${\cal H}$ are given by (39.66, 60.02, 80.03, 99.98, 120.27), 
which justify our choice of LRs and KRs. 
We used static kernel functions which were calculated in advance, 
as explained in the previous section. 
The kernel functions which belong to the 
central KR are compared with the MLWF in Fig. \ref{FIGKRNL}. 

The ground state of this system was calculated 
iteratively by the conjugate gradient method \cite{RCP} with 
no preconditioning. 
Ground state calculations were repeated 100 times 
from different random initial states \cite{FNOTE6}, 
from which statistics were collected. 
Each calculation was terminated successfully when the total energy 
difference between two successive steps was smaller than $10^{-11}$. 
If convergence was not achieved after 1000 iterations, 
the calculation was regarded as a failure, 
which was excluded from the statistics. 

\begin{figure}
  \begin{center}
  \includegraphics[width=160mm]{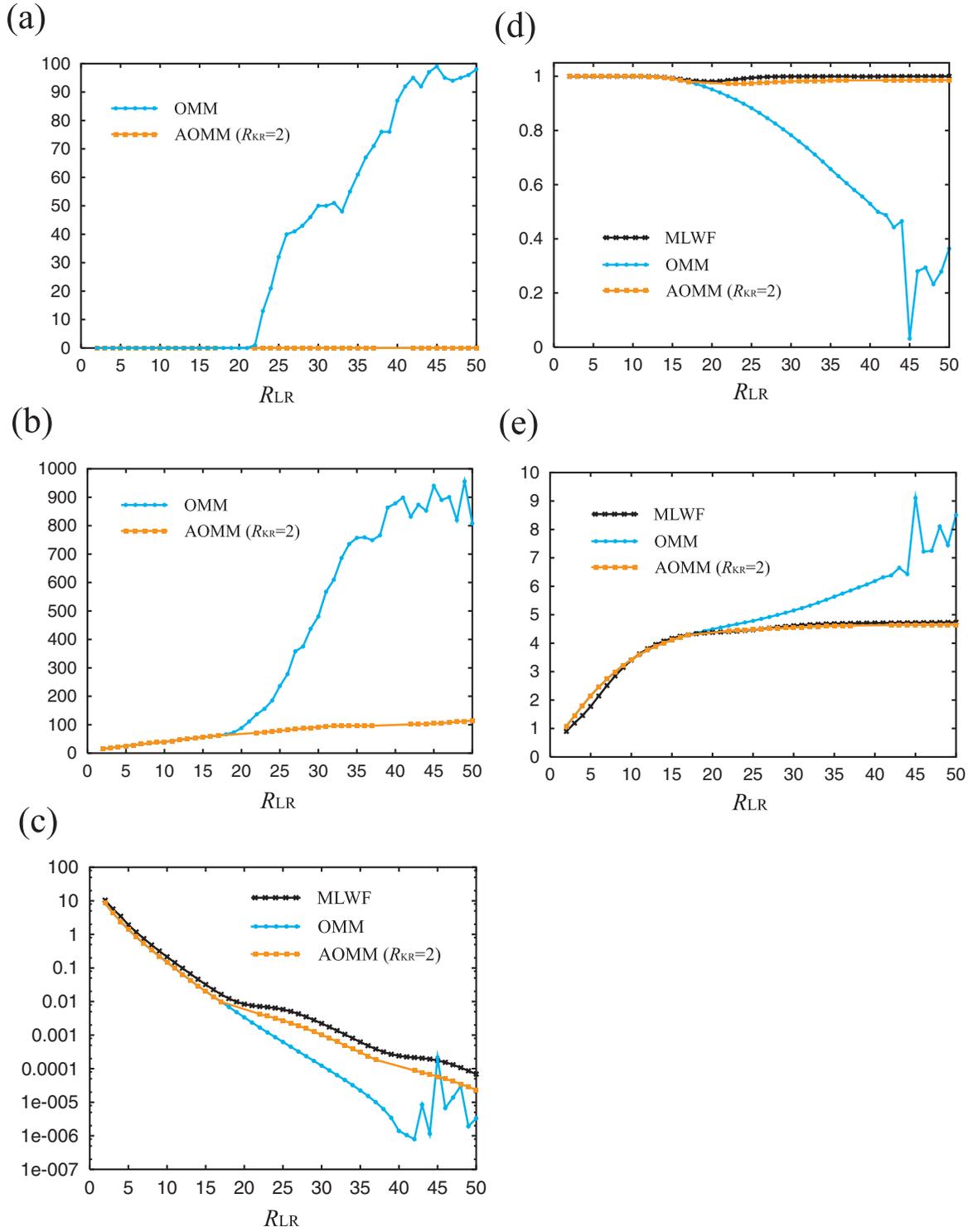}
  \end{center}
  \caption{(a) Number of unsuccessful calculations. 
  	   (b) Average number of iterations. 
	   (c) Relative errors in total energy. 
	   (d) Determinant of the overlap matrix. 
	   (e) Average spread of the orbitals. }
  \label{RES1D}
\end{figure}

Fig.\ref{RES1D} (a) shows the number of unsuccessful calculations 
as functions of $\RLR$ for the OMM and AOMM. 
For small values of $\RLR$, where only a small portion of the 
neighboring LRs overlap, both methods work equally well. 
In the OMM, however, this number grows rapidly 
as the LRs begin to include the centers of 
neighboring LRs at $\RLR \approx 20$, 
and the iterations almost always fail to converge 
when the second nearest neighbors are also included at $\RLR \approx 40$. 
In contrast, no failure is observed in the AOMM for all values of $\RLR$,  
which clearly demonstrates the advantage of AOMM over the OMM. 

Average number of iterations for OMM (Fig.\ref{RES1D} (b)) 
shows a similar tendency. 
While the convergence rate also slowly deteriorates 
with $\RLR$ in the AOMM, 
this problem is easily overcome by a suitable preconditioner 
and/or the multigrid method \cite{BECK00}. 

Fig.\ref{RES1D} (c) shows the relative errors in 
total energy from the exact value obtained by 
diagonalization of ${\cal H}$. 
For comparison, we also show the values for the 
MLWFs, which are first projected onto the LRs of 
size $\RLR$, followed by smoothing at the boundaries. 
While the OMM gives the fastest convergence with respect to $\RLR$, 
the errors saturate at $\RLR \approx 40$, 
because the optimization is always trapped at local minima. 
In contrast, no saturation is observed in the results of AOMM, 
even if the convergence is slower 
due to the additional constraints of Eq.(\ref{LORTHO}). 
Overall, AOMM values are very close to those of MLWFs, 
including the slowdown at $\RLR \approx 20$ and $40$, 
but converge somewhat faster. 
Moreover, no local minima were found in the AOMM. 

The determinant of the overlap matrix $S$ 
at the ground state is shown in Fig.\ref{RES1D} (d). 
While AOMM and MLWF behave similarly, the asymptotic value of 
AOMM ($\approx$ 0.986) is slightly smaller than that of MLWF (=1). 
In contrast, OMM values keep decreasing with $\RLR$, which implies that the 
orbitals are falling into linearly dependent states. 

Fig.\ref{RES1D} (e) shows the average spread $\sigma$ of the orbitals, 
where $\sigma = \sum_{i=1}^N (<\!x^2\!>_i - <\!x\!>_i^2)^\frac12 / N $. 
Although AOMM and MLWF give very similar results, 
OMM values increase steadily with $\RLR$, 
which suggests that the orbitals 
deviate from the picture of MLWFs at large $\RLR$. 

\begin{figure}
  \begin{center}
  \includegraphics[width=90mm]{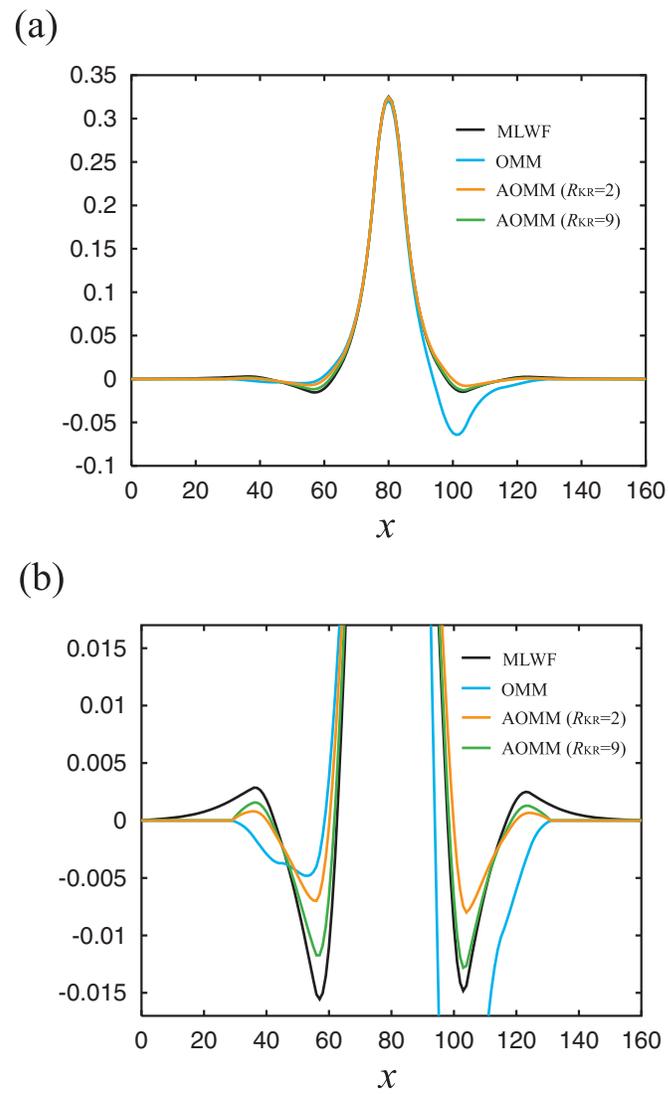}
  \end{center}
  \caption{(a) Localized orbitals which belong 
	to the central LR ($\RLR=50$), obtained from 
	OMM and  AOMM ($\RKRNL$=2 and 9)). 
	(b) Enlarged view of (a).}
  \label{FIGORBS}
\end{figure}

To promote further understanding of this point, 
the optimized orbitals which belong to the central LR 
are compared with the MLWF in Fig.\ref{FIGORBS}. 
A prominent feature of the MLWF is the oscillatory behavior 
at large distances from the center, 
called the orthogonalization tail \cite{QMC05}, 
which arises from the orthogonality constraints of Eq.(\ref{ORTHOG}). 
While the orbitals obtained from AOMM are very similar 
to the MLWF, they decay faster at large distances, 
particularly when $\RKRNL$ is small. 
In contrast, the orbital from OMM exhibits irregular behavior, 
as expected from the large $\sigma$ mentioned above. 

\begin{figure}
  \begin{center}
  \includegraphics[width=95mm]{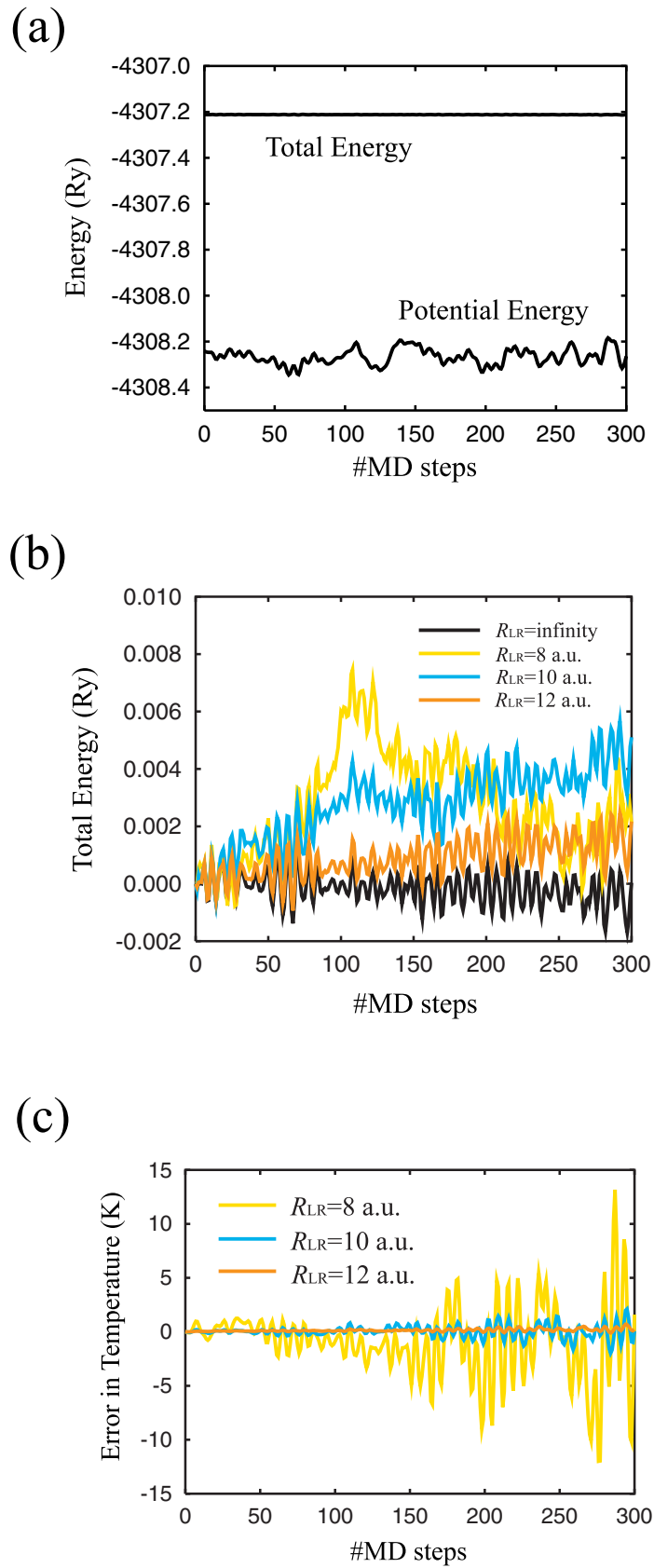}
  \end{center}
  \caption{(a) Time evolution of the total energy and the ionic 
	potential energy when all orbitals are extended. 
	(b) Conservation of the total energy during the simulations. 
	Total energy of the initial configuration is chosen as the origin 
	for each run. 
	(c) Errors in ionic temperature during the simulations. }
  \label{FIGW125}
\end{figure}

We have also implemented AOMM in our first-principles code 
FEMTECK (Finite Element Method based 
Total Energy Calculation Kit) \cite{FEM1,FEM2} 
to assess its performance under realistic conditioins. 
We have carried out ab initio molecular-dynamics simulations 
of liquid water at ambient conditions 
using a cubic supercell of side 29.35 Bohr containing 125 molecules. 
All hydrogen atoms in the system were given the mass of deuterium, 
and a timestep of 40 a.u. ($\sim$ 0.97 fs) was used 
in all simulations. 
We used 125 pairs of LRs and KRs, all of which are centered at the 
oxygen atoms, and 4 orbitals were assigned to each LR and KR. 
The orbitals were optimized using a limited-memory variant 
of the quasi-Newton method \cite{LINO,GILE,QNFEM} 
with a tolerance of $2\times10^{-10}$ Ry/orbital. 
Other details of the simulations are described in our recent publications 
\cite{FA1,PA1}. 
Table \ref{JOBS} shows the details of 4 runs, 
where dynamic kernel functions were used in all AOMM runs \cite{FNOTE5}. 
Fig. \ref{FIGW125}(a) shows the time evolution of the 
total energy and potential energy 
for extended orbitals, which proves the accuracy of the ionic forces 
in our simulations. 
Fig.\ref{FIGW125}(b) and (c) show the total energies and errors in ionic 
temperature during the molecular-dynamics simulations. 
Ionic forces were calculated under the assumption that 
all LRs and KRs are fixed in space. 
In reality, neither assumption is true, 
which explains the irregular behavior of the total energies 
when $\RLR$ is small. 
However, conservation of the total energy for $\RLR=12$ Bohr is 
already competitive with that of the extended orbitals. 
The ionic temperature in the OMM run 
is also reproduced with an error of $<$ 1 K when $\RLR=12$ Bohr. 
\newpage

\section{DISCUSSION}

In this section, 
we make several observations on the properties of AOMM. 

When all LRs are extended, each LR will include all the KRs. 
In this case, 
Eq.(\ref{LORTHO}) imposes $N^2-N$ constraints, which is 
equivalent to the number of degrees of freedom associated with 
Eq.(\ref{XIJ}) (assuming that each orbital is normalized). 
Therefore, the ground state of the system is uniquely determined 
(except for sign) with no loss of accuracy, 
since the redundancy of the orbitals is completely removed. 
If the LRs have a finite size smaller than the unit cell, 
$\etotal$ is no longer invariant under the transformation of 
Eq.(\ref{XIJ}). 
Nevertheless, if a large portion of two LRs overlap with each other, 
$\etotal$ would be nearly invariant under 
the mixing of two orbitals which belong to these LRs. 
This near-redundancy is considered the major source 
of slow convergence and local minima problem \cite{GOED99,BOW00}. 
If these LRs are denoted by $L_1$ and $L_2$, 
we can expect that $K_1 \in L_2$ and $K_2 \in L_1$, 
since the KRs are located near the centers of LRs. 
Then, Eq.(\ref{LORTHO}) gives two constraints on these orbitals, 
which can eliminate the near-redundancy 
associated with $L_1$ and $L_2$. 
On the other hand, if only a small portion of $L_1$ and $L_2$ overlap, 
they do not cause any problems, as shown in the previous section. 
Therefore, the above observation for the extended orbitals 
remains essentially valid even if the orbitals are localized. 

In the limiting case of large (yet nonoverlapping) KRs, 
the static kernel functions will be rather good 
approximations to the MLWFs. 
If the kernel functions are regarded as the zeroth-order approximation to 
the ground state, the orbitals can be written as follows: 
\begin{equation}
|\,\psi_i\!> = |\,\chi_i\!> + |\,\delta\psi_i\!>.  
\end{equation}
Then, the constraints of Eq.(\ref{LORTHO}) would be equivalent to 
\begin{equation}
<\!\chi_j \, | \, \delta\psi_i\!> = 0, 
\end{equation}
which is in close analogy with the case of 
perturbation theory \cite{PUT00,BEN01}. 
Note, however, that our calculations are fully self-consistent. 
On the other hand, if the precise positions of MLWF centers are 
known a priori, e.g. in perfect crystals, the KRs can be chosen 
infinitesimally small, in which case 
each kernel function would be a $\delta$-function. 
Then, Eq.(\ref{LORTHO}) reduces to $\psi_i (\rvec_j) = 0$, 
where $\rvec_j$ denotes the position of $K_j$. 
Since we can expect that $\psi_i (\rvec_i) \ne 0$ for any $i$, 
these constraints will guarantee the linear independence of the orbitals.  
While it may seem counterintuitive, 
the total energy is systematically lower for smaller KRs, 
if each KR is chosen appropriately. 
This is explained as follows. 
When KRs are large, the ground state orbitals 
resemble the conventional MLWFs, 
which suffer from large orthogonalization tails. 
As the KRs become smaller, the influence of Eq.(\ref{LORTHO}) 
becomes more local, thus reducing the 
orthogonalization tails of the orbitals. 
Therefore, from the point of view of minimizing the errors in total energy 
for given LRs, the KRs should be chosen as small as possible. 

So far, we have implicitly assumed that the positions of MLWFs centers, 
which are required for the determination of KRs and LRs, 
are known a priori with sufficient accuracy. 
Fortunately, in many systems with large energy gaps, e.g. in liquid water, 
the electrons form a closed shell. 
Then, approximate positions of MLWF centers are available 
based solely on the knowledge of chemistry. 
If, however, part of the system consists of 
complex atomic configurations with unknown electronic structures, 
it would be difficult to choose the KRs and LRs appropriately. 
One possible solution to this problem is the implementation of 
the adaptive localization centers \cite{FG04,FG06}, 
which gives approximate positions of MLWF centers 
without a priori knowledge of the system. 
An alternative approach is 
to use extended LRs for the orbitals, the behavior of which is 
unpredictable.  
This problem will be discussed in more detail in future publications. 

\newpage

\section{CONCLUSION}

In this article, we have shown that the linear scaling method 
based on OMM can be as robust as the conventional algorithm 
using extended orbitals, 
when augmented with additional constraints to guarantee 
linear independence of the orbitals. 
Although it is difficult to give a general proof, 
AOMM appears to overcome the slow convergence and local minima 
problem of OMM, provided that LRs, KRs, and kernel functions are 
chosen appropriately. 
A more detailed study on the performance of AOMM 
in molecular-dynamics simulations is under way, and 
will be reported in a forthcoming paper.

\section*{ACKNOWLEDGEMENTS}
The author would like to thank K.~Terakura and T.~Ozaki 
for fruitful discussions. 
This work is supported by the Next Generation Supercomputing Project, 
Nanoscience Program, MEXT, Japan.

\newpage

\begin{table}[t]
\caption{Simulation details for each run. 
$M$ and $\Delta E$ denote the average number of iterations and 
the error in total energy for the initial configuration, respectively. }
\label{JOBS}
\begin{center}
\begin{tabular}{cccccc}
\hspace*{1cm} & \hspace{1.5cm} & \hspace{2cm} & \hspace{2cm} & \hspace{2cm} & \hspace{2cm}  \\
\hline
\hline
 & Method &  $\RLR$ (Bohr)  & $\RKRNL$ (Bohr) &  $M$  & $\Delta E$\ (Ry) \\
\hline
(a) &  OMM & $\infty$  &   --     & 12.0  &   0       \\
(b) & AOMM &   8       &   0.8    & 14.1 & 0.13456    \\
(c) & AOMM &  10       &   0.8    & 14.8 & 0.02040    \\
(d) & AOMM &  12       &   0.8    & 14.9 & 0.00309    \\
\hline
\hline
\end{tabular}
\end{center}

\end{table}

\clearpage

\begin{thebibliography}{99}
\bibitem{FNOTE7} No unoccupied states are used throughout the paper.
\bibitem{FNOTE4} We assume that the size of LRs does not 
	increase with system size.
\bibitem{FNOTE1} A small O($N^3$) overhead is 
	required to invert the overlap matrix $S$, 
	which is minor in realistic calculations.
\bibitem{FNOTE2} Of course, a smoother function may be used to 
	avoid discontinuities at the boundaries.
\bibitem{FNOTE6} The initial orbitals were normalized, 
	but not orthogonalized.
\bibitem{FNOTE5} We were able to achieve satisfactory convergence 
	with OMM only when $\RLR=\infty$.
\end{thebibliography}
\end{document}